\documentclass[journal=mamobx,manuscript=article,layout=traditional]{achemso}
\SectionNumbersOn

\usepackage[T1]{fontenc} 
\usepackage{graphicx}
\usepackage{amsmath}
\usepackage{amssymb}
\usepackage{subfloat}
\mciteErrorOnUnknownfalse
\usepackage[dvipsnames]{xcolor}
\usepackage{hyperref}
\hypersetup{        
colorlinks=true,
linkcolor=orange,
citecolor=magenta,
filecolor=green,      
urlcolor=cyan
}

\def\Rg{R_{\mathrm{g}}}
\def\Rg0{R_{\mathrm{g0}}}

\def\Free{\mathcal{H}[z(x)]}
\def\AvRg2{\langle R_g^2 \rangle}
\author{Gabriel Catalini}
\email{gabriel.catalini@uns.edu.ar}
\affiliation{Instituto de Física del Sur (IFISUR), Departamento de Física, Universidad Nacional del Sur (UNS), CONICET,  Av. L. N. Alem 1253, B8000CPB - Bahía Blanca, Argentina}

\author{Nicolás A. García}
\affiliation{Instituto de Física del Sur (IFISUR), Departamento de Física, Universidad Nacional del Sur (UNS), CONICET,  Av. L. N. Alem 1253, B8000CPB - Bahía Blanca, Argentina}

\author{Daniel A. Vega}
\affiliation{Instituto de Física del Sur (IFISUR), Departamento de Física, Universidad Nacional del Sur (UNS), CONICET,  Av. L. N. Alem 1253, B8000CPB - Bahía Blanca, Argentina}

\author{Arash Nikoubashman}
\email{anikouba@ipfdd.de}
\affiliation{Institute of Physics, Johannes Gutenberg University Mainz, Staudingerweg 7, 55128 Mainz, Germany}
\alsoaffiliation{Leibniz-Institut f{\"u}r Polymerforschung Dresden e.V., Hohe Stra{\ss}e 6, 01069 Dresden, Germany}
\alsoaffiliation{Institut f{\"u}r Theoretische Physik, Technische Universit{\"a}t Dresden, 01069 Dresden, Germany}

\title{Simulations of Thin Polymer Films on Flat and Curved Substrates}

\begin{document}


\begin{abstract}
We report molecular dynamics simulation results on the equilibrium properties of polymer thin films adsorbed onto flat and curved substrates. We first systematically determine the contact angle of polymer droplets on flat substrates as a function of the substrate-monomer adsorption strength and degree of polymerization. Focussing on the fully wetted regime, we then investigate the effect of the substrate's topography on the polymer film configuration by using substrates with varying mean curvature. We find that polymer chains close to the substrate are significantly stretched (compressed) in regions of high negative (positive) mean curvature. Further, we find partial dewetting near regions of high mean curvature for thin polymer films. For thicker films, the polymers fully cover the substrate and the shape of the liquid-vapor interface largely follows the shape of the substrate. As the film thickness is increased further, the liquid-vapor interface becomes completely flat. These simulation results are corroborated by analytic calculations using a simplified continuum model that treats the adsorbed polymers as an incompressible medium. Our findings show how the substrate topography influences the conformation of individual chains as well as the shape of the entire film.
\end{abstract}

\section{Introduction}
Polymer thin films have become indispensable in our daily lives, finding applications in a myriad of areas, including food and product packaging \cite{Siracusa2008}, protective coatings \cite{Hyon2018} and lubricants \cite{Adibnia2020}. As technology advances, these films are becoming thinner, approaching the nanometer scale. Such confined systems are generally characterized by large surface to volume ratios, which can lead to strong deviations from the bulk behavior, even at regions far away from the confining interfaces. Consequently, the materials properties of such thin films often differ substantially from their bulk counterparts: For example, interfaces can drastically reduce the structural relaxation dynamics of polymers\cite{priestley:sci:2005, burroughs:mm:2016} or dictate the orientation and morphology of microdomains in block copolymer thin films.\cite{segalman:mserep:2005, darling:progress:2007, hamley:progress:2009, tsarkova:adv:2010, nikoubashman:mm:2013, nikoubashman:mm:2014, huang:pr:2021}

Polymer films can be realized in many varieties, the most common one being that of a thin film supported by a flat substrate. Experimentally, it is relatively simple to create such a film through spin casting. If the substrate is atomically smooth, then it is generally possible to create even polymer films with minimum roughness. The desired film thickness is directly linked to the solution concentration and/or the angular velocity of the spin coater,\cite{Extrand1994, Lawrence1988} and the resulting film thickness can be readily characterized by techniques like X-ray reflectometry \cite{Foster1990} or ellipsometry \cite{Walsh1999}. The film properties might also depend on the specific processing conditions, {\it e.g.}, whether the films are created {\it via} solvent casting, dip coating or spray coating.\cite{tang:epj:2016, mueller:progress:2020} 

From a theoretical point of view, the physical properties of thin polymer films are challenging to predict, since confinement-induced entropic effects\cite{nikoubashman:mm:2014, spencer:mm:2022, matsen:jcp:2022} and interfacial interactions between the chemical constituents can play important roles. In planar films on smooth substrates, the local polymer conformations primarily depend on the distance perpendicular to the substrate, $z$, which simplifies the theoretical description. Films with a thickness $H$ much larger than the radius of gyration of the constituent polymers, $R_\text{g}$, are typically characterized by a thin layer of adsorbed polymers near the substrate, followed by a bulk-like region with chain conformations as in the melt. At the liquid-vapor interface, the monomer concentration usually decays exponentially and the chains exhibit collapsed conformations due to the concomitant decrease in the (effective) solvent quality.\cite{doruker:mm:1998, berressem:jpcm:2021, wang:ml:2023}

While many applications require laterally homogeneous films that fully wet the substrate, it is often observed that films instead spontaneously rupture and thus dewet the substrate. Although extensive research has been conducted on understanding these (de)wetting phenomena on planar substrates,\cite{redon:prl:1991, reiter:prl:1992, seemann:prl:2001, milchev:jcp:2001, becker:nm:2003, Midya2022} the behavior of polymers on substrates with corrugated or chemically patterned topographies is much less understood:\cite{Geoghegan2003} The local curvature and chemistry can have far reaching effects on the conformation of the adsorbed polymers, leading to deviations from classical wetting theories and the emergence of complex interfacial phenomena. However, non-planar substrates also open up new opportunities for controlling the film formation: For example, surface patterns can be used to guide the phase separation of polymer blends \cite{boeltau:nat:1998, karim:pre:1998} or to improve the long-range order of the self-assembled microdomains in block copolymer thin films.\cite{gunkel:small:2018, Vega2013} In general, when a polymer film is constrained to reside on a curved surface, the polymer chains need to compress or stretch to hold the system on the curved geometry. Thus, the curvature induces conformational changes that increase the free energy. The geometric features of a two-dimensional curved surface can be characterized in terms of a shape operator ${\bf S}$, which has two Eigenvalues $\kappa_{1,2}=1/R_{1,2}$ corresponding to the inverse maximal and minimal radii of curvature $R_i$. The determinant and the trace of ${\bf S}$ define the Gaussian curvature and the mean curvature, respectively \cite{Deserno, Vu2018}.

For lamella-forming block copolymers on substrates with shallow periodic (sinusoidal) trenches and zero mean curvature, theoretical considerations suggest that the copolymers should preferentially orient their lamellar domains in the direction perpendicular to the substrate curvature in order to avoid strong chain distortions.\cite{Pickett1993, Pezzutti2015} Man {\it et al.} confirmed these entropic effects through self-consistent field theory (SCFT) calculations, and demonstrated how such wavy substrates can be leveraged to improve long-range ordering.\cite{man:mm:2015, man:mm:2016}. Also using SCFT studies, Vu {\it et al.} analyzed curved films of cylinder-forming block copolymers in cylindrical geometry, finding that the cylinders tend to align along the direction of curvature at high curvatures.\cite{Vu2018} Thus, both experiments and theory show that curvature can be used to manipulate and align nanostructured polymer thin films. Note that while there are several studies concerning the coupling between mean and Gaussian curvature with the textures developed by block copolymer systems, still little is known about the conformational properties of homopolymer systems adsorbed to curved substrates. Evidently, understanding and predicting the behavior of thin polymer films on curved and/or patterned substrates is challenging, as small local variations can have unexpected and long-ranged effects. This situation is somewhat reminiscent of the classic fairy tale ``The Princess on the Pea'', which narrates a prince's quest for an authentic princess through a test involving a hidden pea under multiple mattresses, identifying her by her sensitivity to discomfort.\cite{Princesstale} To gain a better understanding of how local deviations in the substrate flatness impact the structure of polymer thin films, we performed particle-based simulations of homopolymers adsorbed onto substrates with one-dimensional Gaussian shape (locally varying mean curvature and zero Gaussian curvature). This approach allows us to identify the effect of substrate mean curvature on the (local) film thickness and conformation of the constituent chains.

\section{Model and methods}
\label{section:md_simulations}
We performed molecular dynamics (MD) simulations of a generic polymer model to understand the general behavior of polymers adsorbed to flat and curved substrates. The homopolymer chains are represented using the standard Kremer-Grest \cite{Everaers2020} bead-spring model with $M=24$ monomers per chain, unless stated otherwise explicitly. The monomers are modeled as spherical beads with diameter $\sigma$ and mass $m$, which set the units of length and mass of the system, respectively. Non-bonded monomers interact through the standard Lennard-Jones (LJ) pair potential
\begin{equation}
    U_{\rm m}(r) =
    \begin{cases}
    4 \varepsilon \left[\left(\frac{\sigma}{r}\right)^{12} - \left(\frac{\sigma}{r}\right)^6\right], & r \leq r_{\rm cut}
    \\
    0 , & r > r_{\rm cut}
    \end{cases},
    \label{eq:ULJ}
\end{equation}
where $r$ is the distance between a pair of beads, $r_\text{cut} = 3\,\sigma$ is the cut-off distance of the potential, and $\varepsilon$ is the fixed interaction strength between monomers. To ensure that the potential and force go smoothly to zero at the cut-off distance, $U_\text{m}(r)$ was multiplied with a smoothing function
\begin{equation}
    S(r) = \frac{\left(r_{\rm cut}^2 - r^2\right)^2\left(r_{\rm cut}^2 +2r^2 - 3r_{\rm sm}^2\right)}
    {\left(r_{\rm cut}^2-r_{\rm sm}^2\right)^3}.
    \label{eq:ULJ_Sr}
\end{equation}
for distances $r \geq r_{\rm sm} = 2.5\,\sigma$. Polymer bonds were implemented using the finitely extensible nonlinear elastic (FENE) potential\cite{Bishop1979}
\begin{equation}
    U_{\rm b}(r) =
    \begin{cases}
    -\frac{k}{2}r_0^2 \ln\left[1-\left(\frac{r}{r_0}\right)^2\right] , & r < r_0  \\
    \infty , & r \geq r_0
    \end{cases}
    \label{eq:Ufene} .
\end{equation}
with spring constant $k=30\,\varepsilon/\sigma^2$ and maximum bond extension $r_0=1.5\,\sigma$. These parameters prevent unphysical bond crossing \cite{Grest1986}.

The substrates are represented by a Monge parametrization\cite{MongeWolfram} $z(x,y)$ of unidirectional modulation (no Gaussian curvature) of the form:
\begin{align}
    G(x) = A_\text{g} \exp\left(-\frac{x^2} {2 S_\text{g}^2}\right)
    \label{eq:wallShape}
\end{align}
where the parameters $A_\text{g}$ and $S_\text{g}$ control the height and width of the Gaussian bump, respectively.

\begin{figure}[htbp]
    \centering
    \includegraphics[width=8cm]{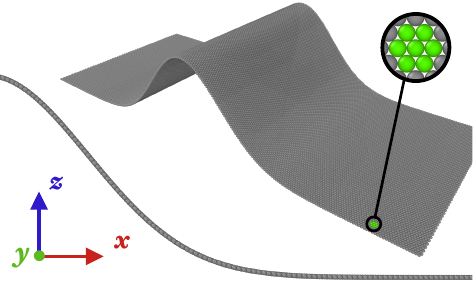}
    \caption{An example of the discretely modeled substrates used in the simulations (surface beads not drawn at scale to improve clarity). The parameters of the Gaussian bump are $A_\text{g} = 20\,\sigma$, $S_\text{g}=10\,\sigma$. Perspective and side view. Some substrate beads are colored green for making the hexagonal array evident.}
    \label{fig:wall_discrete}
\end{figure}

The substrates are modeled by placing discrete particles of diameter $\sigma$ on a hexagonal lattice with lattice constant $0.5\,\sigma$ (see Fig.~\ref{fig:wall_discrete}). Note that the local lattice constant varies slightly (less than $1\,\%$), since the total arc length of the substrate is not always commensurate with the box dimensions. The monomers interact with these wall beads {\it via} the same LJ potential as in Eq.~\eqref{eq:ULJ}, but with a reduced attraction strength $\varepsilon_\text{w} = 0.35\,\varepsilon$, unless specified otherwise explicitly. This parameter is chosen such that the total net interaction leads to fully wetting of the film on the flat substrate (see Sec.~\ref{sec:wetting} below).

The MD simulations have been conducted in the canonical ensemble, unless stated otherwise explicitly, where a Langevin thermostat with friction coefficient $\xi = m/\tau $ was used to fix the temperature to $T= \varepsilon/k_\text{B}$. The equations of motion were integrated using a velocity Verlet algorithm with time step $\Delta t = 0.005\,\tau$, with $\tau = \sqrt{m\sigma^2/(k_{\rm B}T)}$ being the intrinsic MD unit of time. All films have been simulated in a simulation box of dimensions $L_x = 103.6\,\sigma$ and $L_y = 44.21\,\sigma$ with periodic boundary conditions applied to the $x$ and $y$ directions. Polymer films consisted typically of $10^4$ to $10^5$ monomers. All simulations have been performed using the HOOMD-blue software package (v. 2.9.7).\cite{anderson:cms:2020} Visualization and part of the analysis was done using the Open Visualization Tool package.\cite{Stukowski2009}

\subsection{Characterization of the film}
\begin{figure}
\centering
    \includegraphics[width=8cm]{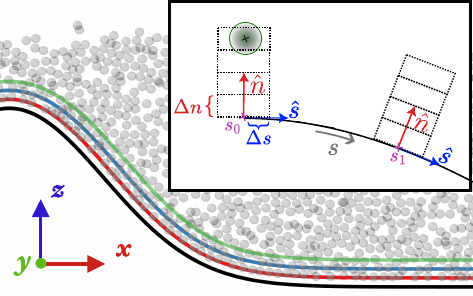}
    \caption{Spatial binning in curvilinear coordinates. The black line represents the substrate, while selected offset curves with constant normal distance from the substrate are shown as colored lines. Some monomers are shown as transparent beads. Inset: Schematic of the approach used to compute properties along the curved substrate.}
    \label{fig:intrinsic_binning}
\end{figure}

Properties of the bulk systems are denoted by the subscript $0$ and are shown in Table S1. For the flat substrates, we characterize the local polymer film properties depending on the distance $z$ perpendicular to the substrate, such as the monomer density $\rho(z)$, the polymer center-of-mass (CM) density $\rho_\text{CM}(z)$, and the mean square radius of gyration of the chains $\AvRg2 (z)$ \cite{Smith2004, Bitsanis1990}. For the curved substrates, we work both with Cartesian 2D maps in the $xz$-plane, and with plots in curvilinear coordinates as described in Fig. \ref{fig:intrinsic_binning}. Here, we explore offset curves \cite{Pham1992} as a function of arc length and conversely analyze fixed positions on the curve as function of normal distance. Each bin is a rectangular cuboid of sides $\Delta n = 0.2\,\sigma$, $\Delta s = 0.5\,\sigma$ and $L_y$.

For reporting chain properties, we use the center-of-mass position of the chain to perform the spatial binning. Due to the nature of the curvilinear coordinates, normal lines on positive (negative) curvature regions will converge (diverge). Thus, a monomer/polymer may contribute to more than one bin in the arc length direction. We averaged the results over $500$ snapshots, separated by $1000\,\tau$, and used the mirror symmetry of the substrate to improve statistics.

\section{Results and Discussion}
\subsection{Adsorption on flat substrates}
\label{sec:wetting}
The stability of a polymer film depends crucially on the wetting conditions between the monomers and the substrate. The basic equation that governs capillary phenomena at interfaces is Young's equations, which relates the interfacial tensions $\gamma_{ij}$ between the substrate (s), liquid (l) and vapor (v) phases to the contact angle $\theta$ that the liquid-vapor interface forms with the solid substrate
\begin{equation}
    \cos(\theta) = \frac{(\gamma_\text{sv} - \gamma_\text{sl})}{\gamma_\text{lv}} = 1 + \frac{S}{\gamma_\text{lv}} ,
    \label{eq:youngs_equation}
\end{equation}
where $S = \gamma_\text{sv} - (\gamma_\text{sl}+\gamma_\text{lv})$ is the so-called spreading coefficient. For a droplet that fully wets the substrate, the contact angle at equilibrium is $\theta= 0^\circ$ such that $S \geq 0$. To form stable thin films, the adhesive substrate-monomer forces must be strong enough to outweigh the cohesive forces in the polymer film. Otherwise, the polymers will form a droplet with a contact angle of $\theta > 0^\circ$ that might even desorb completely ($\theta = 180^\circ$). 

\begin{figure}[htb]
\centering
    \includegraphics[width=8cm]{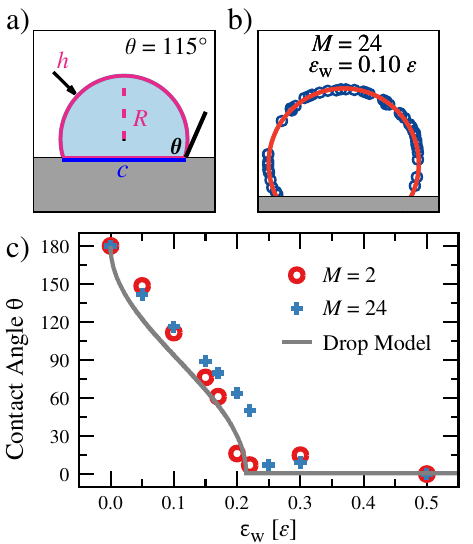} 
    \caption{(a) Droplet model of spherical cap with radius $R$ and contact angle $\theta$. (b) Droplet vapor-liquid interface from MD simulations and its corresponding fitting circle, for chain length $M = 24$. (c) Contact angles $\theta$ as functions of adhesion strength $\varepsilon_{\rm w}$, comparing the results between the model (line) and the adsorption simulations for two different chain lengths, as indicated (symbols).}
    \label{fig:wetting_energies_theta}
\end{figure}

In our simulation model, the adhesion strength is directly controlled through the parameter $\varepsilon_\text{w}$ of the LJ potential acting between monomers and substrate beads, while the strength of the cohesive forces within the film are tuned {\it via} the monomer-monomer interaction strength $\varepsilon$. In the following, we will present a simple continuum model to estimate the contact angle of a (polymer) droplet on a flat substrate as a function of $\varepsilon_{\rm w}$. We assume that the polymers form a spherical cap with radius $R$ and contact angle $\theta$ [Fig.~\ref{fig:wetting_energies_theta}(a)]. We approximate the potential energy of the system (neglecting contributions from bonded interactions, which are identical in all cases) as:
\begin{equation}
    E = \left(V_{\rm p} - \frac{A_{\rm p}h}{2}\right) e_0 \rho_0 + Z A_{\rm srf} h \rho_0\varepsilon_{\rm w}
    \label{eq:Eads}
\end{equation}
where $V_\text{p}$ and $A_\text{p}$ are the volume and surface area of the spherical cap, and $A_{\rm srf} = \pi c^2$ is the surface area of the disk in contact with the substrate. Note that $A_\text{p}$ includes the area of the droplet-substrate interface, $A_{\rm srf}$ [indicated by blue region in Fig.~\ref{fig:wetting_energies_theta}(a)]. Further, we assume a homogeneous monomer number density within the droplet with value identical to the bulk, {\it i.e.}, $\rho_0 \approx 0.91\,\sigma^{-3}$ for our model. From the bulk simulations, we further determined the potential energy per particle due to the LJ potential as $e_0 \equiv E_0/N \approx -3.6\,\varepsilon$ (Table S1). In deriving Eq.~\eqref{eq:Eads}, we have also assumed that the monomers in the outermost droplet layer with thickness $h=\sigma$ have half the number of neighbors, and that the potential energy due to the wall-monomer interaction can be described by a step-like potential with (effective) coordination number $Z=16.88$, that characterizes the number of contacts between an adsorbed monomer and substrate beads. Since the polymer droplet is incompressible, the droplet volume $V_\text{p}$ must be preserved making $R \equiv R(\theta)$. With this constraint, the potential energy written in Eq. \eqref{eq:Eads} becomes a function of the contact angle $\theta$. We then identify the preferred contact angle $\theta$ of our model by finding the minimum of Eq.~\eqref{eq:Eads}.

Figure~\ref{fig:wetting_energies_theta}(c) shows that the contact angle $\theta \to 180^\circ$ in the limit of $\varepsilon_\text{w} \to 0$, as expected. With increasing wall attraction $\varepsilon_\text{w}$, $\theta$ monotonically decreases until it saturates at $\theta = 0^\circ$ for $\varepsilon_\text{w} \gtrsim 0.2\,\varepsilon$. To test these estimates of the contact angle based on simple energetic considerations, we performed MD simulations in which we initially placed a hemispherical polymer droplet of radius $R=30\,\sigma$ (consisting of $N=51456$ monomers) on a flat substrate. We then periodically increased the wall attraction in steps of $\Delta \varepsilon_\text{w}=0.05\,\varepsilon$ and let the system evolve for $1000\,\tau$. We verified that this time is sufficient to let the polymer droplet relax into its equilibrium shape, and determined the contact angle $\theta$ of the droplet with the planar substrate [Fig.\ref{fig:wetting_energies_theta}(b)]. The resulting contact angles $\theta$ from the MD simulations and our continuum model are in excellent agreement, which provides an {\it a posteriori} justification for the simplifications of our model calculation. 

At this point, it should be noted that our continuum model does not consider chain connectivity effects explicitly, but only implicitly through the bulk monomer number density $\rho_0$, which depends on the degree of polymerization $M$.\cite{johnson:jcp:1994} To test the effect of this approximation, we performed additional MD simulations of shorter dimer chains and determined the resulting contact angles $\theta$. As can be seen in Fig.~\ref{fig:wetting_energies_theta}(c), there are only minor differences in $\theta$ for the two different chain lengths: At a given $\varepsilon_\text{w}$, the contact angle $\theta$ was systematically lower for the dimers, and the transition to the fully wetted regime occurred at a weaker wall attraction. These findings are in agreement with recent MD simulations and density functional theory calculations by Midya {\it et al.}\cite{Midya2022} who observed that the temperature for complete wetting ($\theta=0^\circ$) increased with increasing chain length $M$. The differences between the short and long polymers fully disappear for attraction strengths well above the wetting transition, $\varepsilon_\text{w} \gg 0.2\,\varepsilon$. In practice, however, one can not use an arbitrarily large interaction strength $\varepsilon_\text{w}$, because a too strong substrate-monomer attraction will result in an increase of the local monomer density near the substrate, eventually crossing the phase boundary to a crystalline solid. Indeed, for $\varepsilon_\text{w} \gtrsim \varepsilon$ we observed that the monomers near the substrate formed a thin crystalline layer with hexagonal in-plane order, which greatly reduced the mobility of the adsorbed monomers (see SI).

\begin{figure}
\begin{center}
    \includegraphics[width=8cm]{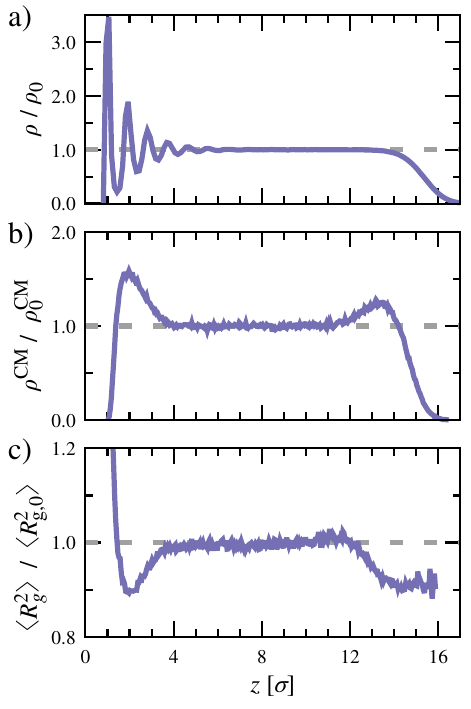}
    \caption{Properties of a thin polymer film with thickness $H \approx 14.8\,\sigma$ deposited on a flat substrate. (a) Local monomer number density $\rho(z)$, (b) local polymer number density $\rho^\text{CM}(z)$, and (c) mean squared radius of gyration $\AvRg2$ {\it vs} distance to the substrate $z$. All data have been normalized with the corresponding values in a bulk melt.}
    \label{fig:flatfilm}
\end{center}
\end{figure}

To establish a reference for polymer films on curved substrates, we first simulated a thin polymer on a planar substrate (see SI for details on the film preparation). We set the substrate-monomer interaction strength to $\varepsilon_\text{w} = 0.35\,\varepsilon$, so that the polymers fully wet the substrate. The nominal film thickness is set to $H \approx 14.8\,\sigma \approx 6.2\,R_{\text{g},0}$, with $R_{\text{g},0} \equiv \left\langle R_{\text{g},0}^2 \right\rangle^{1/2} \approx 2.4\,\sigma$ being the root mean square radius of gyration of the chains in the bulk melt. Figure~\ref{fig:flatfilm}(a) shows the normalized monomer number density in the direction perpendicular to the substrate, $\rho(z)/\rho_0$. Near the substrate, the density profiles show a distinct layering, which is typical for short molecules near hard walls,\cite{snook:jcp:1978} with the first peak occurring near the minimum of the integrated substrate-monomer interaction potential, $z \approx \sigma$ (integrating the 12-6 potential over the discrete lattice results in a 10-4 potential\cite{Bitsanis1990} which has the minimum at $z=\sigma$ instead of $z=2^{1/6}\,\sigma$). The density oscillations diminish with distance from the substrate, and the local monomer density in the thin film approaches the bulk value at about $5$ monomer layers away from the substrate. At the liquid-vapor interface, the monomer concentration decreases rapidly to zero, with $\rho(z)$ following the characteristic $\tanh$ shape. If we consider the location of the polymers based on their center-of-mass positions, we find two local maxima near the substrate-liquid and liquid-vapor interfaces [Fig.~\ref{fig:flatfilm}(b)]. The surplus of chains near the interfaces can be understood by considering the conformation of chains in those regions, which are flattened in the directions tangential to the substrate and shrunk in the direction perpendicular to the substrate [Fig.~\ref{fig:flatfilm}(c)].\cite{doruker:mm:1998, berressem:jpcm:2021}.

\subsection{Polymer films on curved substrates}
\begin{figure}
    \centering
    \includegraphics[width=8cm]{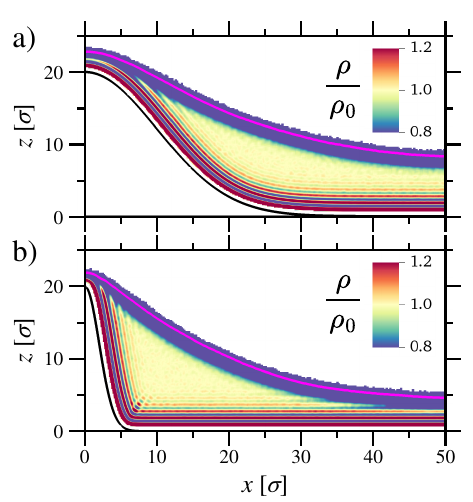}
    \caption{2D colormaps of the normalized monomer density for films with thickness $H=7.4\,\sigma$ on substrates of height $A_\text{g}=20\,\sigma$ and widths (a) $S_\text{g}=10\,\sigma$ and (b) $S_\text{g}=2\,\sigma$. The liquid-vapor interface, as determined by $\rho/\rho_0=0.1$, is shown as a magenta line in the density plots. The colorbars are capped at $\rho/\rho_0 = 0.8$ and $\rho/\rho_0 = 1.2$ to enhance contrast in the colormap.}
\label{fig:colormaps_rho}
\end{figure}

Polymer films on the curved substrated are created by placing the end-configurations of the flat film simulations such that the bottom film surface lies $\sigma$ above the top of the Gaussian bump. This distance is within the attraction range of the substrate-monomer interaction (see Sec.~\ref{section:md_simulations}), so that the film slowly adsorbs to the curved substrate. We then let the systems relax for $5.5 \times 10^5\,\tau $, which is typically long enough so that the films do not evolve anymore in time. Snapshots of the deposition process are shown in Fig. S2. Density distributions of fully deposited films are shown in Fig.~\ref{fig:colormaps_rho} as colormaps in the Cartesian $xz$-plane, for films of nominal thickness $H=7.4\,\sigma$ adsorbed on substrates of height $A_\text{g}=20\,\sigma$ and widths $S_\text{g} = 10\,\sigma$ and $S_\text{g}= 2\,\sigma$, respectively. Similar to the flat films, the monomer density in these films is identical to the bulk density in regions sufficiently far away from the interfaces (yellow areas in Fig.~\ref{fig:colormaps_rho}). In a region within $\approx 5\,\sigma$ of the substrate, we observe the typical density oscillations (layering) \cite{Smith2004, Bitsanis1990}. As is the case for flat films at the liquid-vapor interface, we also observe the characteristic decaying profile of the density in the direction perpendicular to the interface. In this region, we define the liquid-vapor interface as the region $\rho/\rho_0=0.1$ (magenta lines in Fig.~\ref{fig:colormaps_rho}).

\begin{figure}
    \centering
    \includegraphics[width=8cm]{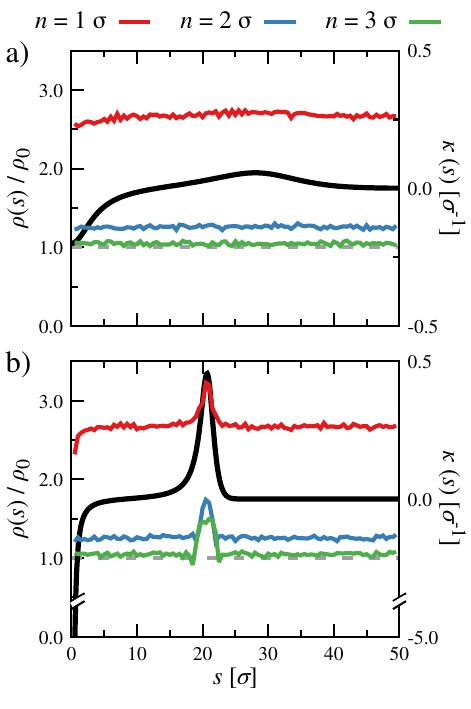}
    \caption{Normalized density $\rho/\rho^0$ (left $y$-axis) for films of thickness $H=22.2\,\sigma$ {\it vs} arc length $s$ for offset curves at distances $n = 1$, $2$, and $3\,\sigma$ (red, blue and green lines, respectively). Substrate of height $A_\text{g}=20\,\sigma$ and widths (a) $S_\text{g}=10\,\sigma$ and (b) $S_\text{g}=2\,\sigma$. The curvature (right $y$-axis) of each substrate is presented as a black solid line. Note that for panel b) the curvature at $s=0\,\sigma$ takes the value $\kappa = -5\,\sigma^{-1}$.}
\label{fig:offsetcurves_density}
\end{figure}

To study the adsorbed polymer layer near the substrate in more detail, we plotted in Fig.~\ref{fig:offsetcurves_density} offset curves of $\rho$ at distances $n = 1$, $2$, and $3\,\sigma$ as a function of arc length coordinate $s$,\cite{Pham1992} as defined in Fig.~\ref{fig:intrinsic_binning}. These curves demonstrate that lines of constant distance are also lines of (almost) iso-density. However, for the narrowest substrate, we observe a peak in density near $s= 20\,\sigma$, where the substrate has the highest curvature $\kappa = 0.47\,\sigma^{-1}$ (lowest positive radius of curvature $R \equiv \kappa^{-1} \approx 2.12\,\sigma$). Thus, the tendency for monomers to wet the substrate induces a local increase in monomer density near the region of highest positive substrate curvature. This effect can be understood by considering the local distribution of substrate monomers in the vicinity of positive curvature ($\kappa > 0$): A monomer in those regions interacts with more surface beads within the cutoff distance $r_\text{c}$ of the attractive substrate-monomer interaction potential, which thus results in a local increase of the monomer density. Similarly, regions of negative curvature ($\kappa < 0$) will experience the inverse effect with a decreased interaction when compared to a flat substrate. 

\begin{figure}[htbp]
    \centering
    \includegraphics[width=8cm]{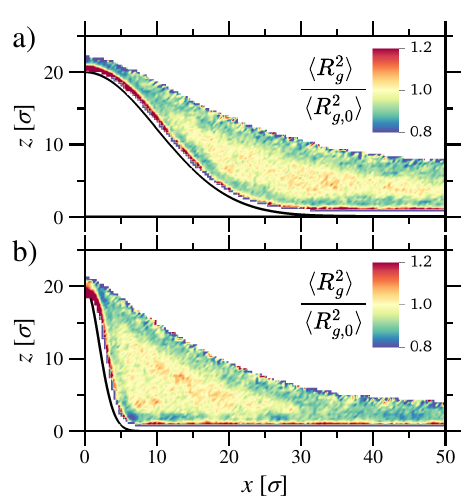}
    \caption{2D colormaps of the normalized mean square radius of gyration, $\langle R_\text{g}^2 \rangle/\langle R_\text{g,0}^2 \rangle$, for films with thickness $H=7.4\,\sigma$ on substrates of height $A_\text{g}=20\,\sigma$ and widths (a) $S_\text{g}=10\,\sigma$ and (b) $S_\text{g}=2\,\sigma$. The colorbars are capped at $\langle R_\text{g}^2 \rangle/\langle R_\text{g,0}^2 \rangle = 0.8$ and $\langle R_\text{g}^2 \rangle/\langle R_\text{g,0}^2 \rangle = 1.2$ to enhance contrast in the colormap.}
\label{fig:colormaps_rg}
\end{figure}

The result for the radius of gyration colormaps are shown in Fig.~\ref{fig:colormaps_rg} for the same parameters as in Fig.~ \ref{fig:colormaps_rho}. We observe that in a region of $n \approx \sigma$ away from the substrate, the chains are stretched along the surface, resulting in a greater $R_\text{g}$ (thin red line in the colormaps). As for the flat films, we also observe that immediately after this layer we encounter a layer of reduced chain size  (blue region). Again, we encounter a region of bulk properties between the substrate-liquid and liquid-vapor interfaces (yellow region). Consistent with the result presented above for the narrowest bump, we observe a dip in the size of the chains. Because we assign the chain property to its center-of-mass position, some bins inside the substrate are filled when chains are stretched along the top of the bump [red curved region in Fig. \ref{fig:colormaps_rg}(b)]. From these data, it appears that regions of high positive (negative) curvature induce shrinking (stretching) of the adsorbed chains.

\begin{figure}
    \centering
    \includegraphics[width=8cm]{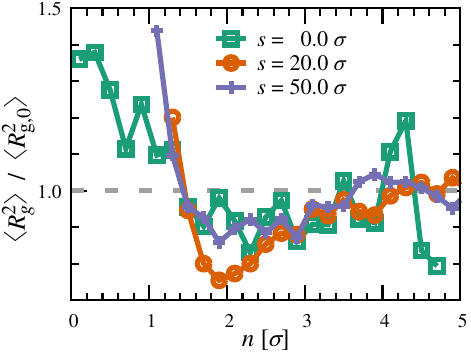}
    \caption{Normalized radius of gyration $\langle R_g^2 \rangle/\langle R_{g,0}^2 \rangle$ {\it vs} normal distance $n$ for a film of thickness $H=22.2\,\sigma$ adsorbed onto substrate of height $A_\text{g}=20\,\sigma$ and width $S_\text{g}=2\,\sigma$. Data shown at positions $s = 0\,\sigma$ (green squares, $\kappa = -5.0\,\sigma^{-1}$) , $s = 20\,\sigma$ (orange circles, $\kappa = 0.47\,\sigma^{-1}$) and $s = 50\,\sigma$ (purple crosses, $\kappa = 0\,\sigma^{-1}$).}
\label{fig:tangentialcurves_rg2}
\end{figure}

To support this finding, we plot in Fig.~\ref{fig:tangentialcurves_rg2} the normalized radius of gyration as function of normal distance $n$ at some selected positions $s$. Here, we focus on a film of thickness $H=22.2\,\sigma$, such that we have full coverage on the top of the bump. Again, we find that the chains are stretched in regions $n \approx \sigma$ away from the substrate, whereas chains in regions of positive curvature (orange circles in Fig.~\ref{fig:tangentialcurves_rg2}) are significantly compressed when compared to chains near the flat regions of the surface (purple crosses in Fig.~\ref{fig:tangentialcurves_rg2}). Likewise, near the top of the substrate, where the curvature is negative (green squares in Fig.~\ref{fig:tangentialcurves_rg2}), we can find stretched chains even for regions of $n<\sigma$, since the the center of mass of these chains can fall in regions within the substrate.

For all substrates, we found that the liquid-vapor interface becomes flatter with increasing film thickness $H$, which indicates that polymers far away from the substrate do not experience the curvature anymore. To better understand (and possibly predict) the shape of the adsorbed polymer films, we constructed a continuum model, where the polymer film is treated as a homogeneous incompressible medium that fully wets the substrate. Within this framework, the free energy of the system is given by:

\begin{equation}
    \Free = L_y \int_{-L_x/2}^{L_x/2} \gamma_\text{lv} \sqrt{1 + (\nabla z(x))^2}\text{d}x
    \label{eq:hamiltonian_interface}
\end{equation}
where $z(x)$ denotes the position of the liquid-vapor interface. This expression corresponds to the surface energy of the modulation $z(x)$ relative to a flat plane which is a smooth function (no terraces) of $x$. Thus, the liquid-vapor interface $z(x)$ of the deposited film can be computed by solving the variational problem of minimum surface (arc length in 1D) for the film with constraints (i) $L_y \int \left[z(x)-G(x) \right] \text{d}x = V$ to preserve the total film volume (incompressibility), and (ii) $z(x) \geq G(x)+h_\text{min}$ (full wetting of the film). These conditions are imposed by penalty factors in the free energy functional:
\begin{equation}
    \frac{\mu}{2} \int_{-L_x/2}^{L_x/2} \left[z(x)-G(x)-V \right]^2 \text{d}x
\end{equation}
and
\begin{equation}
    \frac{\nu}{2} \Theta \left[z(x)-G(x)-h_\text{min} \right]^2 ,
\end{equation}
where $\Theta$ is the Heavisde function, and $\mu, \nu$ are the penalization parameters for the fixed volume and minimum thickness conditions, respectively. Since the underlying substrate $G(x)$ is a symmetric function of $x$, the solution will be symmetric also.

\begin{figure}[htbp]
\centering
\includegraphics[width=8cm]{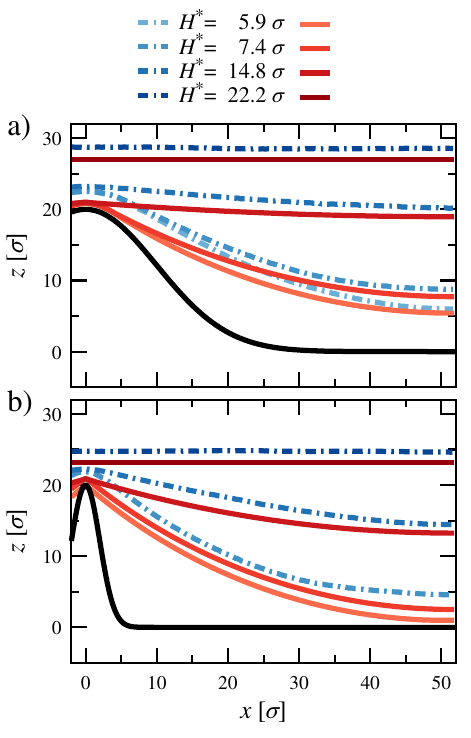}
\caption{Comparison of liquid-vapor interfaces obtained from the MD simulations (dashed blue lines) and the continuum model (solid red lines) for substrates of height $A_\text{g}=20\,\sigma$ and widths (a) $S_\text{g}=10\,\sigma$ and (b) $S_\text{g}=2\,\sigma$.}
\label{fig:lv_interfaces}
\end{figure}

We convert the problem to its Weak Form, which allows greater flexibility in handling complex boundary conditions and constraints and solve it on an one-dimensional grid of spacing $\Delta x = 0.1\,\sigma$ (roughly $\approx 1000$ points) with no-flux boundary conditions at $x=-L/2$ and $x=L/2$. Table S2 specifies the parameters used in the numerical solution. Figure \ref{fig:lv_interfaces} shows the comparison of the liquid-vapor interfaces obtained from MD simulations (blue dashed lines) and from solving the variational problem (red solid lines). We compare values of equal film thickness, assuming a constant monomer number density for the MD simulations $H^{*} = N/(\rho_0 L_x L_y$). Note that local variations in the monomer density $\rho(z)$ [Fig.~\ref{fig:colormaps_rho}(a)] lead to a small mismatch between $H^*$ from simulations and our continuum calculations. Further, it is important to mention that polymer films with $H^* = 5.9\,\sigma$ were unstable in the MD simulations with a narrow bump $S_\text{g} = 2\,\sigma$, as the polymers dewetted the region of high curvature near the bump. In a similar manner, for the continuum model, enforcing the minimum thickness condition was not possible for this film thickness, resulting in a liquid-vapor interface that crossed the solid substrate. For the rest of the cases, the parameter $\nu$ was chosen (see Table S3) such that the solution converged and the condition was met. The resulting shapes of the liquid-vapor interface from MD simulations and the continuum model are in qualitative agreement: As the film thickness increases, the information from the surface modulation is increasingly lost, until the liquid-vapor interface becomes completely flat for sufficiently large thicknesses. 

\section{Conclusions}
In this work, we have studied the adsorption of thin homopolymer films onto flat and curved substrates, using a combination of particle-based simulations and continuum-level theory. For the flat substrates, we have systematically characterized the wetting regimes by determining the equilibrium contact angle $\theta$ as a function of adsorption strength and degree of polymerization. Here, we found that $\theta$ decreased monotonically with increasing adsorption strength, until the polymers fully covered the substrate. We further found that shorter chains require a stronger substrate-monomer attraction to achieve the same wetting behavior as longer chains. For flat substrates in the fully wetted regime, we observed characteristic density oscillations of the monomers for approximately 5 layers, which then transitioned to the bulk value and decayed exponentially at the liquid-vapor interface. The distribution of the polymer center-of-mass positions showed peaks of similar intensities and widths at the interfaces, which we attribute to the flattened conformation of the chains in those regions. 

For polymers adsorbed to curved substrates, we observed partial dewetting of very thin polymer films in proximity to areas with strong curvature, at conditions where the flat films still remained stable. As the film thickness increased, the polymers fully coat the substrate, and the configuration of the liquid-vapor interface closely mirrors the substrate's shape. Ultimately, the liquid-vapor interface transitions to a completely flat state. The monomer number density remained almost constant when moving along the tangential coordinate at a fixed normal distance. However, we found that the monomer density has a peak in regions of high positive curvature on the substrate. In addition, the polymers seem to stretch at the top of the bump and compress in regions of high positive curvature. 

The results from our particle-based simulations and the simplified continuum model were in decent agreement for the majority of cases, thus providing a computationally cheap method for quickly exploring parameter space. In future work, we will explore how the substrate geometry can be leveraged to direct the order in more complex systems, such as polymer blends or block copolymers.

\begin{acknowledgement}
This work was funded by the Deutsche Forschungsgemeinschaft (DFG, German Research Foundation) through Project Nos. 233630050, 248882694, 405552959 and 470113688.
\end{acknowledgement}
\bibliography{references}

\end{document}


\clearpage
\newpage
\section{Limit $\varepsilon_\text{w} \gg \varepsilon $}
\label{section:hexagonal_order}

To characterize the in-plane ordering of monomers near the substrate, we compute the per particle hexagonal orientational order parameter \cite{Chakrabarti1998}:

\begin{equation}
    \Psi_{6}^{m} = {\frac{1}{N_n}} 
    \sum_{n=1}^{N_n}{e^{6i \theta_{nm}}} .
    \label{eq:phi6m}
\end{equation}

\begin{figure}[htbp]
\centering
\includegraphics[width=8cm]{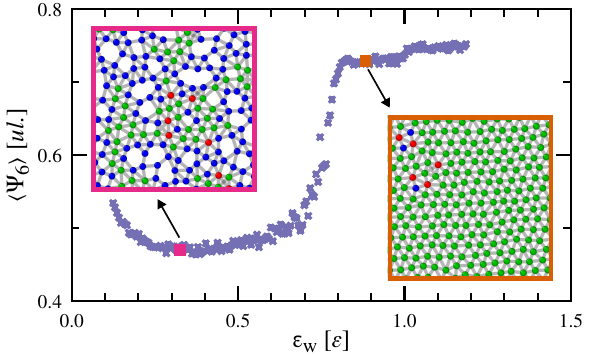}
\caption{Average orientational order parameter $\Psi_6$ as function of wall interaction strength $\varepsilon_\text{w}$, for a polymer film deposited onto a flat substrate. The order parameter is computed for the first layer of adsorbed monomers. Insets: Monomers in the first layer colored by their number of neighbors, namely 5 (blue), 6 (green) and 7 (red).}
\label{fig:Q6_snap}
\end{figure}

For sufficiently strong wall interactions $\varepsilon_\text{w} \gg \varepsilon$, we find hexagonal order in the first layer of adsorbed monomers (Fig.~\ref{fig:Q6_snap}).

\section{Bulk Simulations}

In order to characterize the effect of film confinement, we performed additional bulk simulations in the $NPT$ ensemble, with $N_\text{p} = 2000$ chains at $P=0\,\varepsilon/\sigma^3$ with $\tau_T=0.5\,\tau$ and $\tau_P=1.0\,\tau$. The resulting equilibrium monomer number density is $\rho_0 = 0.91\,\sigma^{-3}$. To investigate the bulk chain properties, we prepared a new $NVT$ simulation with edge length $L_x = L_y = L_z = 56\,\sigma$ containing $N_\text{p} = 6658$ chains. These simulations were run for $150000\,\tau$ for equilibration, and then we took $1000$ samples every $50\,\tau$. A summary of the resulting bulk properties is given in Table~\ref{tab:bulk_props}.

\begin{table*}
\begin{tabular}{clc}
    \hline
    \textbf{Var} & \textbf{Meaning} & \textbf{Value} \\
    \hline
    \hline
    $\rho_0$ & Monomer number density at $P=0$ & $0.91\,\sigma^{-3}$ \\
    $e_0$ & Per particle average LJ energy & $-3.6\,\varepsilon$ \\
    $e_\text{bond}$ & Per bond average FENE energy & $18.1\,\varepsilon$ \\
    $ \langle R_{g,0}^2 \rangle $ & Mean square radius of gyration & $5.82\,\sigma^2 $ \\
    $ \langle R_{e,0}^2 \rangle $ & Mean square end-to-end distance & $35.0\,\sigma^2 $ \\
    $ \langle \Delta \rangle_0 $ & Mean asphericity & $0.41$ \\
    $ D_0 $ & Diffusion coefficient from MSD of chains & $1.4 \times 10^{-3}\,\sigma^2/\tau $ \\
    $ \tau_0 $ & End-to-end vector autocorrelation time & $1.1 \times 10^{3}\,\tau$ \\
    \hline
\caption{Table with average values of polymer properties in the bulk.}
\label{tab:bulk_props}
\end{tabular}
\end{table*}

\section{Film Preparation}
To prepare the polymer films, we first generate the desired number of chains as a random walk of fixed bond length $\sigma$ in a box of the desired $L_x$ and $L_y$ dimensions, but with a $L_z$ such that the bead density is low at $ \rho = 0.1\,\sigma^{-3}$. Then, we shrink the box in the $z$ direction during $5000\,\tau$ until the desired bulk density $\rho_0$ is reached. At this point, we unwrap the polymers in the $z$ direction and place them in a much taller ($L_z$) box. We let the system relax for $10^4\,\tau$, where it will form a free standing film. Once a stable configuration has been reached, we measured the surface tension as the difference in pressure in the components normal and tangential to the film according to the standard definition \cite{Kirkwood1949}:

\begin{equation}
    \gamma = \frac{L_z}{2} \big\langle P_{zz} - \frac{P_{xx}+P_{yy}}{2} \big\rangle
    \label{eq:surface_tension}
\end{equation}

\begin{figure}[htbp]
\centering
\includegraphics[width=7cm]{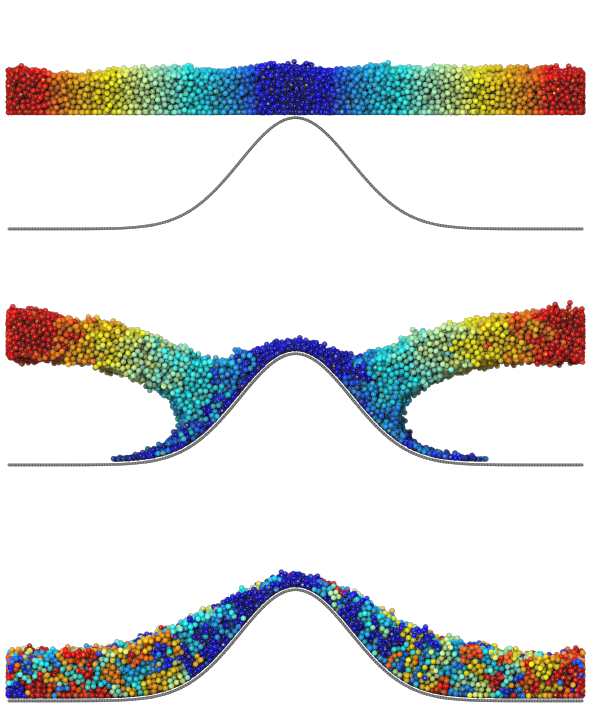}
\caption{Snapshots of MD simulations during the deposition of a film of thickness $H=7.4\,\sigma$ onto a substrate of parameters $A_\text{g}=20\,\sigma$ and $S_\text{g}=10\,\sigma$. Monomers are colored by their initial absolute $x$ position.}
\label{fig:film_deposition}
\end{figure}

For all free standing films, we find a constant surface tension of $\gamma_\text{lv} = 1.235\,\varepsilon\sigma^{-2}$. We then deposit the film on the substrate by placing the film one monomer diameter above the substrate. We let the films adsorb onto the substrates, and relax them for $5.5 \times 10^5\,\tau$, taking samples every $5 \times 10^2\,\tau$. Snapshots of the deposition simulations are shown in Fig. \ref{fig:film_deposition}.

The case where $A_\text{g} = 0\,\sigma $ serves as a check of the results for flat walls. From these flat films, we compute the total surface tension $\gamma_\text{tot}$ of the system {\it via} Eq.~\eqref{eq:surface_tension} and infer the substrate-liquid interfacial tension as $ \gamma_\text{sl}  = \gamma_\text{tot} - \gamma_\text{lv}$. We find a substrate-liquid surface tension of $\gamma_\text{sl} = -2.42\,\varepsilon/\sigma^2$, meaning that the monomers have a strong preference to be adsorbed on the substrate.

\section{Variational Problem}
\label{section:variational_problem}
Table \ref{tab:variational_params} specifies the parameters used in the numerical solution of Eqs.~(7-9) in the main manuscript, subject to the specified constraints. In order to enforce the minimum thickness condition, different values for $\nu$ needed to be used depending on the film thickness and substrate width. Table~\ref{tab:nu_values} summarizes these values along with the maximum height of the film obtained. If the value of $H_\text{max}$ lies below $A_\text{g} + h_\text{min}$ ($= 21\,\sigma$ for all investigated cases), then the minimum thickness condition was not fully achieved.

\begin{table*}
\begin{tabular}{clc}
    \hline
    \textbf{Var} & \textbf{Meaning} & \textbf{Value} \\
    \hline
    \hline
    $\Delta x$ & Grid spacing & $0.1\,\sigma$ \\
    $l=L_x/2$ & Box size & $51.8\,\sigma$ \\
    $H$ & Film thickness & $5-20\,\sigma$ \\
    $h_\text{min}$ & Minimum thickness & $\sigma$ \\
    $\mu$ & Penalty for constant volume & $1$ \\
    $\nu$ & Penalty for minimum thickness & $0.3$ \\
    $A_\text{g}$ & Height of Gaussian bump & $0-20\,\sigma$  \\
    $S_\text{g}$ & Width of Gaussian bump & $2-10\,\sigma$ \\
    \hline
\end{tabular}
\caption{Parameters used for computing the liquid-vapor interface through the variational approach.}
\label{tab:variational_params}
\end{table*}

\begin{table*}
\begin{tabular}{ccccc}
    \hline
    ~ & \multicolumn{2}{c}{$S_\text{g}=2\,\sigma$} & \multicolumn{2}{c}{$S_\text{g}=10\,\sigma$} \\
    \textbf{$H$} & $\nu$ & $H_\text{max}$ & $\nu$ & $H_\text{max}$ \\
    \hline
    \hline
    $5.9\,\sigma$ & 0.9 & $19.7\,\sigma$ & 10 & $\ge A_\text{g} + h_\text{min}$ \\
    $7.4\,\sigma$ & 33 & $20.9\,\sigma$ & 1 & $\ge A_\text{g} + h_\text{min}$ \\
    $14.8\,\sigma$ & 9 & $20.8\,\sigma$ & 30 & $\ge A_\text{g} + h_\text{min}$ \\
    $22.2\,\sigma$ & 0.1 & $\ge A_\text{g} + h_\text{min}$ & 0.1 & $\ge A_\text{g} + h_\text{min}$ \\
    \hline
\end{tabular}
\caption{Values of the parameter $\nu$ for enforcing the minimum thickness condition. In all cases, $A_\text{g} + h_\text{min} = 21\,\sigma$.}
\label{tab:nu_values}
\end{table*}

\clearpage
\newpage
\bibliography{references}